\documentclass[]{interact}

\usepackage{epstopdf}
\usepackage[caption=false]{subfig}

\usepackage[numbers,sort&compress]{natbib}
\bibpunct[, ]{[}{]}{,}{n}{,}{,}

\begin{document}


\title{Magnetization and magneto-transport measurements on CeBi single crystals}

\author{
\name{Brinda Kuthanazhi, Na Hyun Jo\textsuperscript{$^\ast$}\thanks{$^\ast$Current address: Advanced Light Source, Lawrence Berkeley National Laboratory, Berkeley, CA 94720, USA}, Li Xiang\textsuperscript{$^\dagger$}\thanks{$^\dagger$Current address: National High Magnetic Field Laboratory, Florida State University, Tallahassee, Florida 32310, USA
}, Sergey L. Bud'ko, Paul C. Canfield \thanks{CONTACT Author. Email:canfield@ameslab.gov}}
\affil{Ames Laboratory and Department of Physics and Astronomy, Iowa State University, Ames, Iowa-50011, USA}
}

\maketitle

\begin{abstract}
We report the synthesis of CeBi single crystals out of Bi self flux and a systematic study of the magnetic and transport properties with varying temperature and applied magnetic fields. From these $R(T,H)$ and $M(T,H)$ data we could assemble the field-temperature ($H-T$) phase diagram for CeBi and visualize the three dimensional $M-T-H$ surface. In the phase diagram, we identify regions with well defined magnetization values, and identify a new phase region. The magnetoresistance (MR) in the low temperature regime shows, above $6~$T a power-law, non-saturated behavior with large MR ($\sim 3\times10^5 \%$ at $2~$K and $13.95~$T), along with Shubnikov-de Haas oscillations. With increasing temperatures, MR decreases, and then becomes negative for $T\gtrsim 10~$K. This crossover in MR seems to be unrelated to any specific magnetic or metamagnetic transitions, but rather is associated with changing from a low-temperature normal metal regime with little or no scattering from the Ce$^{3+}$ moments and an anomalously large MR, to an increased scattering from local Ce moments and a negative MR as temperature increases.
\end{abstract}


\section{Introduction}

In recent years there has been a revived interest in the properties of low carrier density or semi-metallic materials with large spin-orbit-coupling (SOC).  When these semi-metals are rare earth bearing, the interaction between local moments and conduction electrons can lead to remarkable magnetic field-temperature ($H-T$) phase diagrams as well as dramatic changes in magnetoresistance across multiple phase transitions \cite{Budko1998, Myers1999, Petrovic2002, Petrovic2003, Xiang2019}.  In the $R$Bi, $R$Sb,  $R$Bi$_2$, $R$Sb$_2$, $R$AgBi$_2$, $R$AgSb$_2$ ($R$ = rare-earth) and related systems the sometimes large magnetoresistance, combined with recent computational predictions or models, has lead to interest in possible topological states existing in compounds that can exhibit long range magnetic order \cite{Kuroda2018, Petrovic2012, Petrovic2013, Shi2016}. 

The rare-earth monopnictides ($RX$, $R$=rare-earth, $X$=P, As, Sb, Bi) family, which crystallize in the simple NaCl structure, is such a set of compounds. They have been studied in the past for the rich magnetism they host, including perspective of Kondo lattices and valence fluctuation compounds \cite{Tsuchida1965,Tsuchida1967,Nereson1971,Cable1972,Halg1982,Kasuya1996,Wiener2000}. In recent years new studies have been done on them as candidates for topological materials \cite{Niu2016,Wu2016,Zeng2016,Guo2016,Lou2017,Wu2017,Ye2018,Li2018,Kuroda2018,Ryu2020}. An interesting aspect, that has been brought to the forefront by recent studies is the existence of large magnetoresistance in many of the rare-earth antimonides and bismuthides \cite{Zeng2016,Ye2018,Liang2018,Vashist2019}.

CeBi, the heaviest member of the Ce$X$ family, is known to show multiple metamagnetic transitions in the magnetic field - temperature phase space, with a strong anisotropy arising from crystalline electric field (CEF) effects and a resultant anisotropic magnetoresistance \cite{Bartholin1974,Bartholin1979,RossatMignod1980,RossatMignod1983,Kohgi2000,Lyu2019}. It thus provides an exemplary candidate to study the effects of magnetic ordering in magnetoresistance and electronic structure.

Here, we report the crystal growth, magnetic characterization, and detailed magneto-transport measurements on single crystals of CeBi. We identify regions that are prone to large hysteresis in field as well as temperature, correlating with the existence of many, near degenerate magnetic states. The $H-T$ phase diagram obtained from decreasing temperature and field sweeps reveals an additional phase line as compared to the most complete previous report, but otherwise agrees well with it \cite{Bartholin1979}. We also observed a non-monotonic behavior of magnetoresistance with temperature, where a low temperature large MR behavior slowly diminishes and becomes negative MR with increasing temperatures. This was followed by a shift to positive MR again but with comparatively small magnitude.

\section{Experimental details}

Single crystals of CeBi were grown out of Bi self flux \cite{Canfield1992}, using the current binary Bi-Ce phase diagram \cite{Okomoto2000}. An initial composition of Ce$_{26}$Bi$_{74}$ was sealed in a welded tantalum tube under an argon atmosphere, followed by sealing under partial pressure of argon into a fused silica ampoule. The thus prepared ampoule was heated up to $1200 ^\circ$C over $8$ hours and held there for $4$ hours, followed by slow cooling to $940 ^\circ$C over $60$ hours, at which temperature, the excess flux was decanted using a centrifuge \cite{Canfield1992,Canfield2019}. 
The air sensitive crystals, of typical dimensions of about $5~$mm $\times 5~$mm $\times 3~$mm, were handled in a nitrogen filled glove box. The crystal structure as well as phase purity were confirmed by powder x-ray diffraction using a Rigaku Miniflex diffractometer (also in nitrogen glove box), using Cu $K_\alpha$ radiation. 

\begin{figure}
\centering
\includegraphics[scale=1]{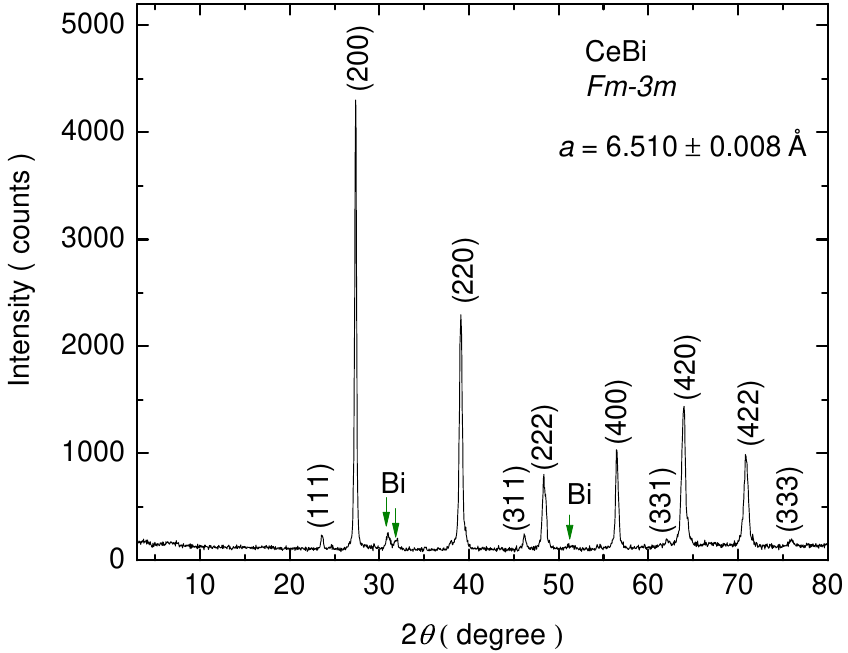}
\caption{Powder x-ray diffraction pattern of CeBi. Peaks are identified and ($hkl$) indices are assigned by comparison with the reported cubic structure of CeBi in space group $Fm\bar3m$ \cite{Yoshihara1975}. Additional peaks due to Bi flux are marked by green arrows. The obtained lattice parameter $a=6.510\pm0.008~$\textrm{\AA} is in agreement with Ref. \cite{Yoshihara1975}.}

\label{pxrd}
\end{figure}

Magnetization ($M$), as a function of temperature ($T$) was measured on two different crystals. An initial measurement of $M(T)$ at $0.5$ T was made in a Quantum Design Magnetic Property Measurement System (MPMS), upon cooling from $300$ K to $2$ K. This is shown below in Fig. \ref{mtrt} (a). Later more detailed and systematic measurements were carried out in an MPMS 3, in vibrating sample magnetometer (VSM) mode, from temperatures $1.8$ K to $35$ K, and applied magnetic fields ($H$) up to $7$ T. The magnetic field was applied along one of the $<001>$ directions. For $M(T)$ data acquisition, measurements were done with both zero field cooled (ZFC) sample with increasing temperatures and on field cooled (FC) with decreasing temperatures. Similarly, for $M(H)$ measurements, both increasing and decreasing field isotherms were measured for various temperatures between $3$ K and $27.5$ K.

Resistance measurements were done in the standard four-probe geometry, in a Quantum Design Physical Property Measurement System (PPMS) using a $1~$mA excitation with a frequency of $17~$Hz. Electrical contacts were made on a cleaved, rectangular bar shaped, crystal, using silver paint (DuPont 4929N). The magnetic field, up to $14~$T, was applied along one of the $<001>$ directions, and perpendicular to the current which was also along one of the $<001>$ directions. For all the measurements, crystals were oriented based on their very clear, natural, cubic morphology and square facets. $R(T)$ measurements were done partially on warming and partially on cooling.  $R(T)$ curves at $0$, 0.5, 1.5, 2, 3.5, 4.2, 4.5, 5.5, 7, 9, and $13.95~$T were obtained by measuring resistance on cooling the sample, and the rest while warming up. Similarly $R(H)$ at 3.5, 4, 7, 10, 12, 20, 25, 35, 40, and $50~$K were measured with decreasing magnetic field, and the rest with increasing fields. 

\section{Results}

Figure \ref{pxrd} shows the powder x-ray diffraction data. The peaks resolved match with the reported peak positions for the cubic $Fm\bar3m$ structure of CeBi \cite{Yoshihara1975}. From the ($hkl$) indices assigned, by comparison to the reported structure, we calculated the lattice parameter $a=6.510\pm0.008~$\textrm{\AA}, which is in agreement with Ref. \cite{Yoshihara1975}. Weak additional peaks, corresponding to small amounts of residual Bi flux are also identified (marked by green arrows in the figure).
\subsection{Magnetization}

\begin{figure}
\centering
\includegraphics[scale=0.55]{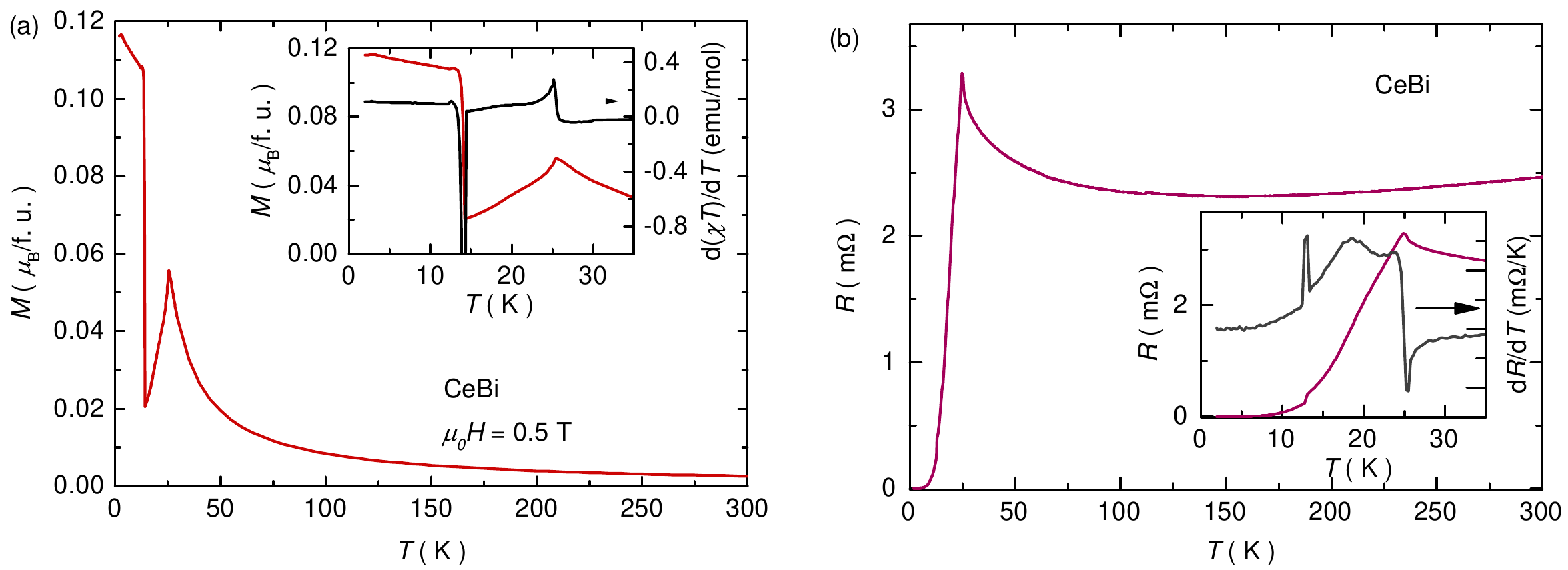}
\caption{(a) Magnetization ($M$) as function of temperature ($T$) for an applied magnetic field of $0.5~$T. Inset: $M(T)$ in the low temperature region from $2~$K to $35~$K, along with the derivative d$(\chi T)$/d$T$ which is the criteria used here for determination of transition temperature. Two transitions around $25~$K and a little below $15~$K are are clearly seen. (b) Resistance ($R$) as a function of $T$ at zero field. Inset: $R(T)$ and d$R$/d$T$ from $2~$K to $35~$K, clearly showing the two transitions at around $25~$K and $12.5~$K. Both $M(T)$ and $R(T)$ data were taken upon cooling.}
\label{mtrt}
\end{figure}

\begin{figure}
\centering
\includegraphics[scale=0.8]{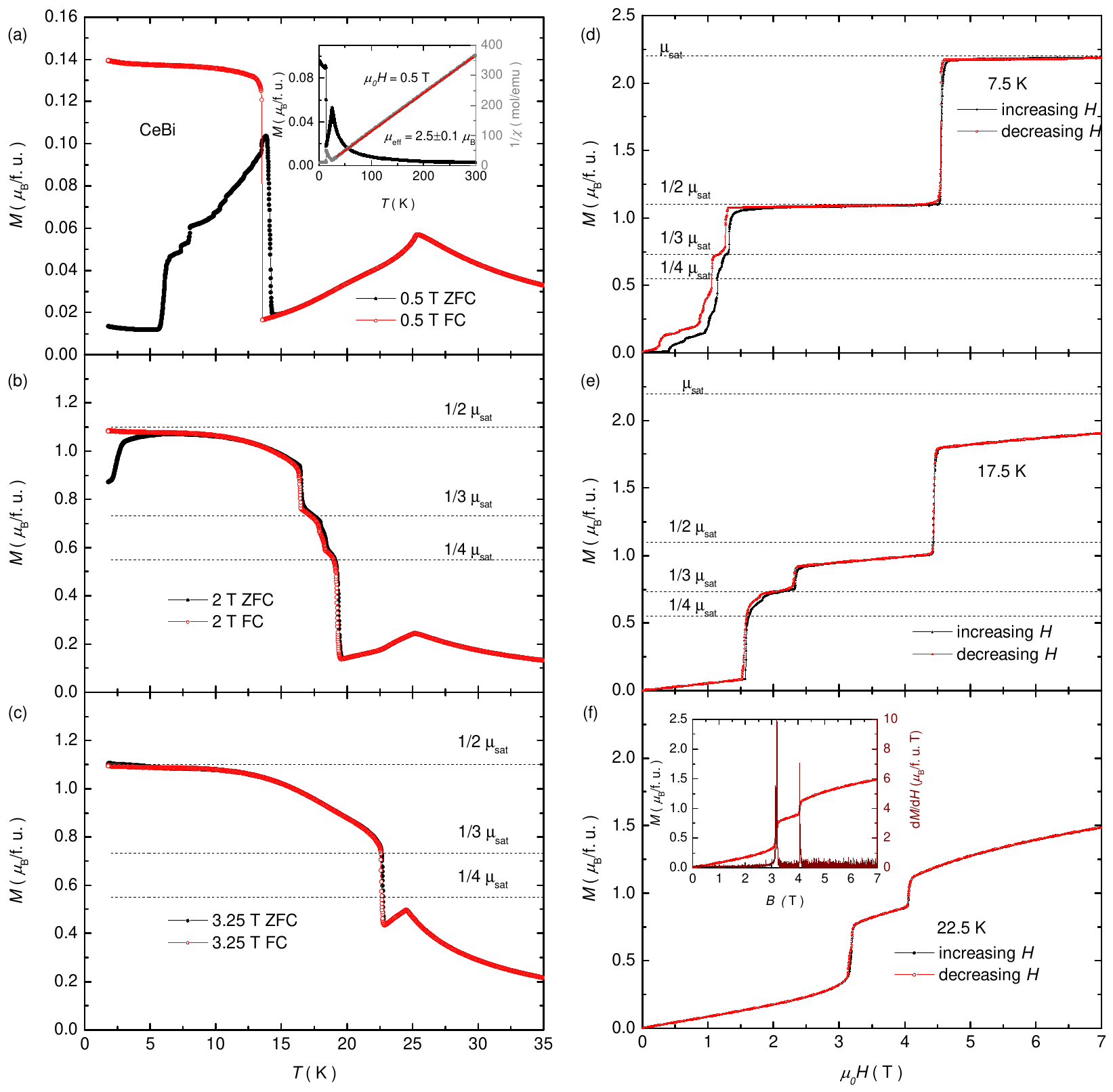}
\caption{Representative sets of $M(T)$ and $M(H)$ data. (a) $M(T)$ at a relatively low field of $0.5~$T measured zero field cooled and field cooled. A clear hysteresis is seen between the two. Inset: A Curie-Weiss fit to the inverse susceptibility data above $50~$K. (b) $M(T)$ measured at an intermediate field value of $2~$T. Except for lowest temperatures, ZFC and FC data overlap. The dotted horizontal lines denote values corresponding to $1/2$, $1/3$, and $1/4$ of $\mu_{sat}$, saturated magnetic moment. (c) $M(T)$ measured at $3.25~$T where ZFC and FC are overlapping. (d) $M(H)$ at $7.5~$K with increasing and decreasing magnetic fields. Below $1.5~$T there is a series of metamagnetic transitions which show a clear hysteresis. Dotted lines show $1/2$, $1/3$, and $1/4~\mu_{sat}$ and $\mu_{sat}$ values. (e) $M(H)$ at an intermediate temperature of $17.5~$K, where the curves almost overlap. (f) $M(H)$ at $22.5~$K where the hysteresis has disappeared. Inset: $M(H)$ plotted along with $dM/dH$. The maxima in $dM/dH$ are used as the criteria to determine the transitions from $M-H$ data. }
\label{mhys}
\end{figure}

The magnetic characterization was done on cleaved and clean crystals handled in a nitrogen filled glove box. Figure \ref{mtrt} (a) shows $M(T)$ data taken on cooling from $300~$K to $2~$K. The inset shows the $2-35~$K range of the $M(T)$ curve, clearly showing the two transitions, a kink-like anomaly at $\sim 25~$K (indicated in Fig. \ref{pd1} below by a black circle) and a step-like increase of $M$  a little below $15~$K (indicated in Fig. \ref{pd1} below by a dark yellow square). The transition temperatures are determined by taking the extrema in $d(\chi T)/dT$ data, which is the criterion for transition temperature in a simple antiferromagnet \cite{Fisher1962}, but has been used to identify multiple transitions as well \cite{Ribeiro2003}. Both from the shape of the derivative curves and from the hysteretic behavior shown in Fig. \ref{mhys}, we can identify that the transition around $25~$K is a second order transition, and the one below $15~$K is first order. This is in agreement with the reported nature of the magnetic transitions \cite{Halg1982}. A Curie-Weiss fit for the paramagnetic regime above the transition, as shown in the inset of Fig. \ref{mhys}(a), gives an effective moment, $\mu_{eff}=2.5\pm0.1~\mu_B$, in agreement with the expected effective moment $2.54~\mu_B$ for Ce$^{3+}$ ions, and $\theta = 10.6\pm0.1~$K,.

Figure \ref{mhys} shows a representative set of $M(T)$ and $M(H)$ data. $M(T)$, measured on both ZFC and FC sample with warming and cooling respectively; data for a low field of $0.5~$T, at an intermediate field of $2~$T, and at a higher field of $3.25~$T are shown in Fig. \ref{mhys} (a), (b) and (c). For lower field measurement shown in Fig. \ref{mhys} (a), there is a stark distinction between ZFC and FC data, with a notable hysteresis for the transition below $15~$K. In addition, the ZFC data go through a cascade of features, some of which are barely resolvable, whereas the FC measurement shows two well defined features. The features in ZFC data are possibly due to a realignment of domains or may be associated with many closely spaced, near degenerate, ordered states. With increasing fields, the differences in ZFC and FC $M(T)$ data start to disappear and eventually they fall on top of each other, as shown for $M(T)$ at $2~$T and $3.25~$T in Fig. \ref{mhys} (b) and (c) respectively. 

\begin{figure}
\centering
\includegraphics[scale=0.6]{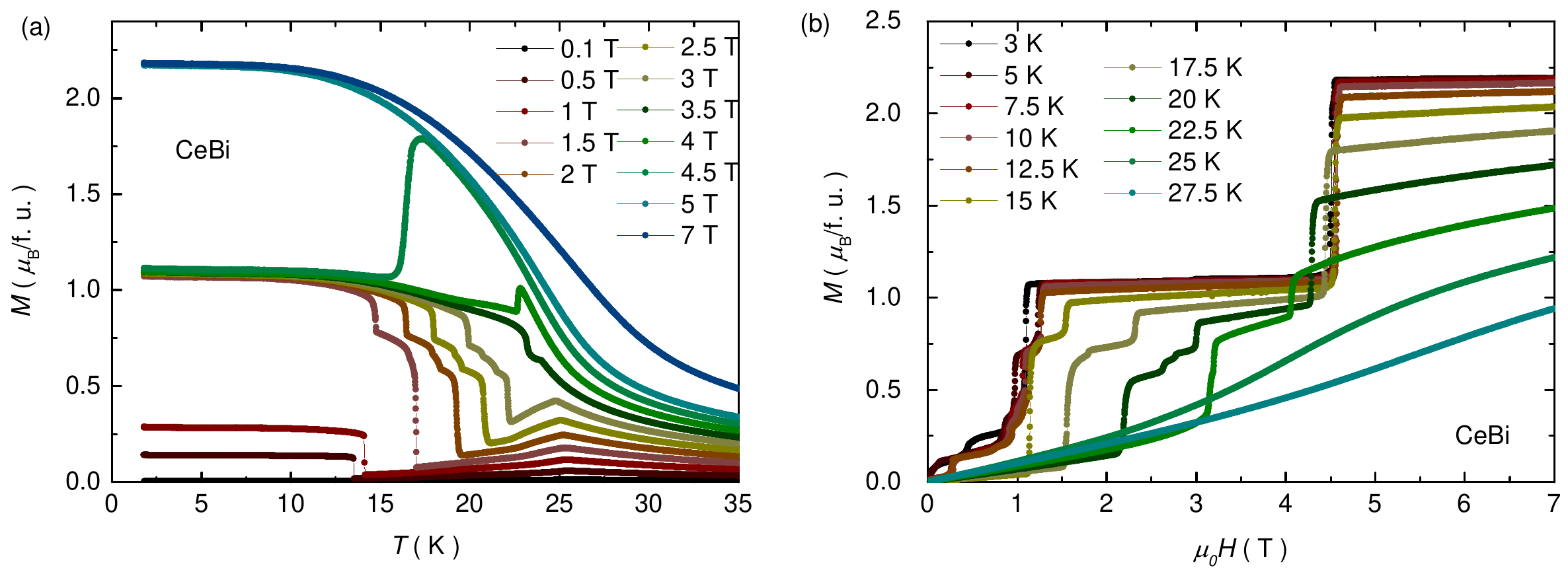}
\caption{(a) Magnetization as function of temperature, measured with decreasing temperature (FC), at various applied magnetic fields ($\mu_0H$) from $0.1~$T to $7~$T. (b) $M$ as a function of $\mu_0H$ measured at various temperatures from $3~$K to $27.5~$K. All data were taken with decreasing field.}
\label{mtmh}
\end{figure}

\begin{figure}
\centering
\includegraphics[scale=0.35]{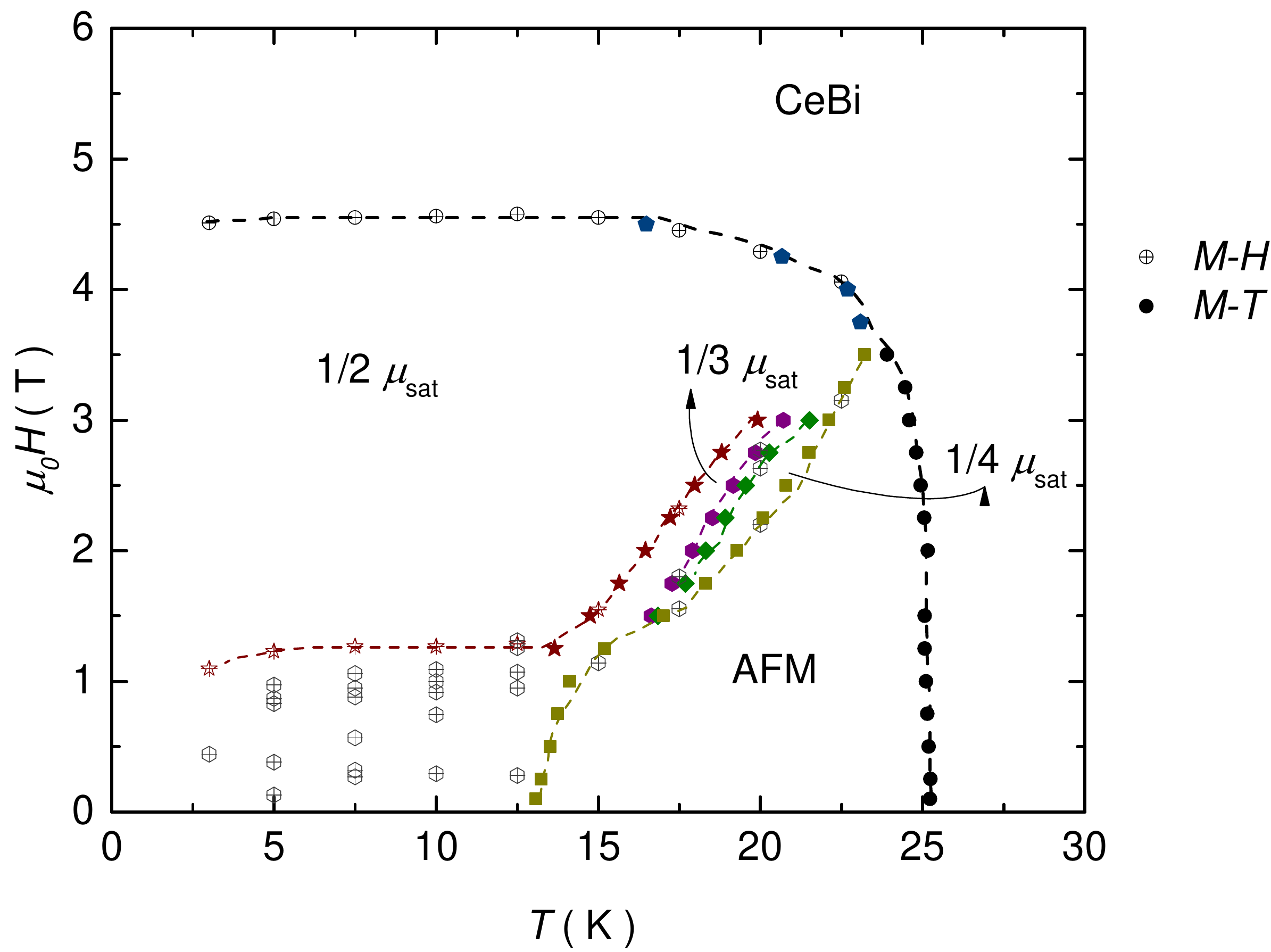}
\caption{Field - temperature ($H-T$) phase diagram of CeBi obtained from the magnetic measurements. Solid symbols and open-crossed symbols are obtained from $M$ as a function of $T$ and $H$ data respectively. Lines are guides to the eye. Regions associated with antiferromagnetic (AFM) order as well as plateaus of approximately $\frac{1}{2}, \frac{1}{3}$, and $\frac{1}{4}$ $\mu_{sat}$ are labelled. The data in the lower temperature, lower field, hysteretic region is not labelled, as the phase boundaries are not clear because of differences in increasing and decreasing field data. Transition temperature, field points in this regime are shown in grey. All the points are determined from FC $M(T)$ and decreasing field $M(H)$ measurements. }
\label{pd1}
\end{figure}

\begin{figure}
\centering
\includegraphics[scale=0.5]{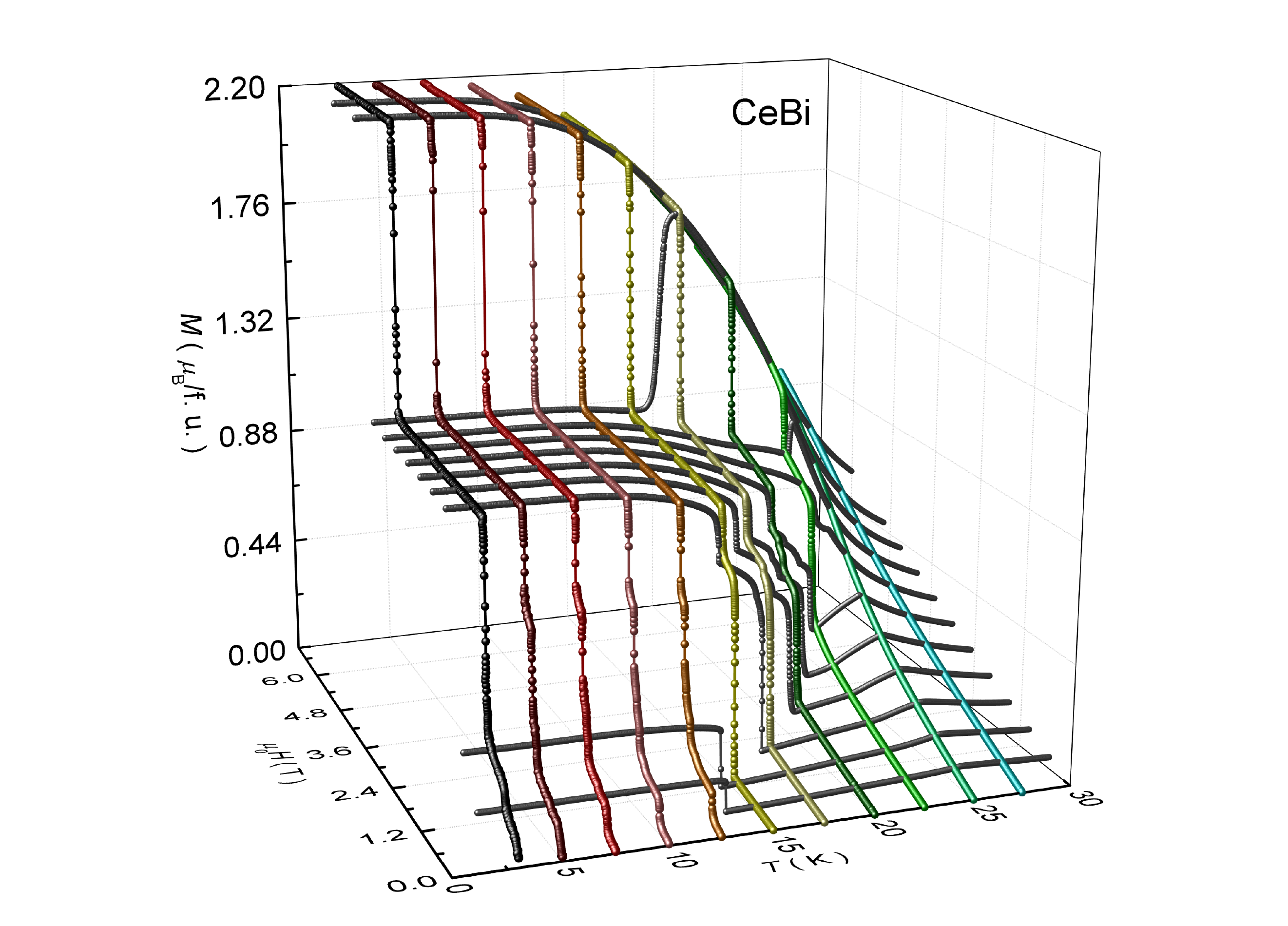}
\caption{Magnetization as a function of both temperature and magnetic field combining Fig. \ref{mtmh} (a) and (b) showcasing various metamagnetic transitions and plateaus. All data are from FC $M(T)$ and decreasing field $M(H)$ measurements. }
\label{mth}
\end{figure}

Similarly, for $M(H)$ data a comparison between a low temperature ($7.5~$K), an intermediate temperature ($17.5~$K) and a higher temperature ($22.5~$K) measurement is shown in Fig. \ref{mhys} (d), (e) and (f). At low temperatures, there exists a series of metamagnetic transitions in the low field regime, with a clear hysteresis, as shown for $M(H)$ at $7.5~$K. The two, clear, higher field metamagnetic states have locally saturated magnetizations of $\mu_{sat}=2.2~\mu_B/f. u. $ (as compared to $2.14~\mu_B$ for Ce$^{3+}$) and approximately $1/2 \mu_{sat} = 1.1~\mu_B$. It should be noted that Fig. \ref{mhys} (b) also shows the plateaus in $M(T)$ corresponds to roughly a quarter, a third, and a half of the measured saturated magnetic moment value $\mu_{sat}=2.2~\mu_B/f. u. $, which are denoted by dotted lines in the figure. As we increase the temperature this lower-field hysteretic behavior starts to disappear and the increasing- and decreasing- field $M(H)$ data fall on top of each other with well defined plateaus, as shown in Fig. \ref{mhys} (e) and (f) for $17.5~$K and $22.5~$K respectively. The inset of Fig. \ref{mhys} (f) shows both $M(H)$ and $dM/dH$ plotted together. The peaks in $dM/dH$ is identified as the transition fields and are used in plotting the phase diagram shown in Fig. \ref{pd1}, which will be discussed later. From here on, all the magnetization data shown, both in Fig. \ref{mtmh} and Fig. \ref{mth}, as well as used for determining the phase diagram in Fig. \ref{pd1} are from FC $M(T)$ and decreasing field $M(H)$ measurements. 

Having gained some insight into the hysteretic behavior of the magnetization data, we can now look at how various features evolve with temperature and field. Figure \ref{mtmh} (a) shows the temperature dependent magnetization, $M(T)$, at various magnetic fields up to $7~$T. As compared to the $0.5~$T measurement, by increasing magnetic fields up to $3.5~$T, the kink-like anomaly near $25~$K becomes more pronounced and is gradually suppressed to lower temperatures. Whereas, the step-like anomaly, initially a little below $15~$K, moves to higher temperature with increasing fields. At about $3.5~$T, these two anomalies merge together and evolve into a single, jump-like drop of $M$ for  $4~$T $\leq \mu_0H \leq 4.5~$T. At even higher fields, all anomalies are suppressed and behavior of $M(T)$ approaches that of a field polarized (saturated paramagnetic) state. In the intermediate field range ($1.5~$T $< \mu_0H < 3~$T), additional step-like anomalies in $M (T)$ appear between $15~$K and $22~$K (corresponding to the series of closely placed transitions shown in Fig. \ref{pd1} using star, hexagon, and diamond symbols) which can be associated with additional magnetic transitions. From these transitions, we can start to build aa $H-T$ phase diagram as shown in Fig. \ref{pd1}. These transition temperatures obtained from $M(T)$ data are shown in Fig. \ref{pd1} as filled symbols (circles, squares, diamonds, and stars).

More of the CeBi $H-T$ phase diagram can be inferred from $M(H)$ data. Multiple magnetic transitions can also be observed in the field dependent magnetization $M(H)$ data as shown in Figs. \ref{mhys} (d), \ref{mhys} (e), \ref{mhys} (f), and \ref{mtmh} (b). At $3$ K, the $M(H)$ curve shows two sharp jumps at $\sim4.5~$T and $\sim 1~$T, and another smaller kink or shoulder at around $0.5~$T. At $3~$K, and $7~$T a saturated moment, $\mu_{sat}=2.2\pm0.1~\mu_B$/f. u. is obtained which is within the error bar from the expected $2.14~\mu_B$ for Ce$^{3+}$ ions. With increasing temperatures, the higher field transition barely moves (e. g. At $7.5~$K, as shown in Fig. \ref{mhys} (d), the higher field transition occurs at $4.5~$T), but below about $1.5~$T one can see a cascade of closely spaced transitions setting in, which are hysteretic in increasing and decreasing field measurements, as can be seen in Fig. \ref{mhys} (d). For $15~$K and above, i.e. above the zero field value of the lower transition temperature, the low field transitions disappear and  $M(H)$ curves become simpler, with well defined plateaus (Figs. \ref{mhys} (e) and (f)). Above $25~$K all the features corresponding to the various transitions disappear, and the $M(H)$ curve resembles that of a typical local moment system in the paramagnetic state. The points for the phase diagram are obtained from $M(H)$ data by evaluating the peaks in derivative $dM/dH$, and are denoted as open-crossed symbols in Fig. \ref{pd1}. 

A field-temperature phase diagram obtained from our $M(T)$ and $M(H)$ data is shown in Fig. \ref{pd1}. This is very similar to the previously reported phase diagrams \cite{Bartholin1974,Bartholin1979,Iwata1991}. A very early study on polycrystalline CeBi was already able to identify three distinct phase regions in the $H-T$ phase space, which were identified as two antiferromagnetic orderings, and a ferrimagnetic ordering with $M=1/2\mu_{sat}$, in addition to the field polarized and paramagnetic regimes \cite{Tsuchida1967}. Later, more detailed measurements on single crystalline samples were reported, with as many as seven different phases in the same $H-T$ regime \cite{Bartholin1974}. They also observed the hysteretic nature of the transitions, which decreased with increasing temperatures. Followed by this, a neutron scattering study confirmed the AFM ordering and partially assigned magnetic structures to the various phases \cite{Bartholin1979}. They identified the two AFM orderings, the $M=1/2\mu_{sat}$ phase and a variety of mixed phases in between, at low temperatures. For $T\gtrsim 12~$K, they identified two phases, but were not able to assign a net magnetization value. Later, a molecular field model, with Ce moments in [001] direction and having an oscillatory exchange interaction perpendicular to the (001) plane, was shown to calculate the phase diagram agreeing very well with the experimental data \cite{Iwata1991}. 

With this information in hand, we can try to interpret the phase diagram we have obtained. The various features in Fig. \ref{pd1} could be understood as follows: Existence of an envelope denoting paramagnetic to antiferromagnetic at higher temperatures, and a full saturation into field polarized state at lower temperatures, with phase boundaries agreeing very well with the existing reports, within which one can see multiple other regions. For $T$ above $15~$K, one can see a well defined region of antiferromagnetic order, and two narrow regimes corresponding to roughly $\frac{1}{4}$ and $\frac{1}{3}$ $\mu_{sat}$ values. There also exist a narrow phase in between these, which was not identified in the previous reports \cite{Bartholin1974,Bartholin1979}. We were not able to assign a locally saturated magnetization value to this phase, given its very limited extent. In the lower temperature regime, between $1.5~$T and $4.5~$T we have an extended region which corresponds to $M=\frac{1}{2}M_{sat}$, which agrees well with the previous reports, whereas, at lower fields, we have plethora of transitions with ill defined and hysteretic phase boundaries, once again agreeing well with reported mixed phases from the neutron study \cite{Bartholin1979}. Nevertheless, we were not able to assign any specific net magnetization values to phases here either, as opposed to Ref. \cite{Bartholin1979}, because of the large hysteresis and irreversible nature of the increasing and decreasing field data.  

Magnetization both as a function of temperature and magnetic field are plotted in Fig. \ref{mth} by combining the data presented in Figs. \ref{mtmh} (a) and (b). As shown in the figure, data from two sets of measurements agree with each other very well and together they depict various magnetic transitions and plateaus. Figures \ref{mhys}, \ref{mtmh}, and \ref{mth} also emphasize that whereas for low temperatures the magnetic plateaus are well saturated at well defined values, as temperature increases, the plateaus-like regions have (i) decreasing extent, (ii) develop finite slopes, and (iii) have their values decrease from their lowest temperature values.

\subsection{Resistance}

\begin{figure}[h!]
\centering
\includegraphics[scale=1]{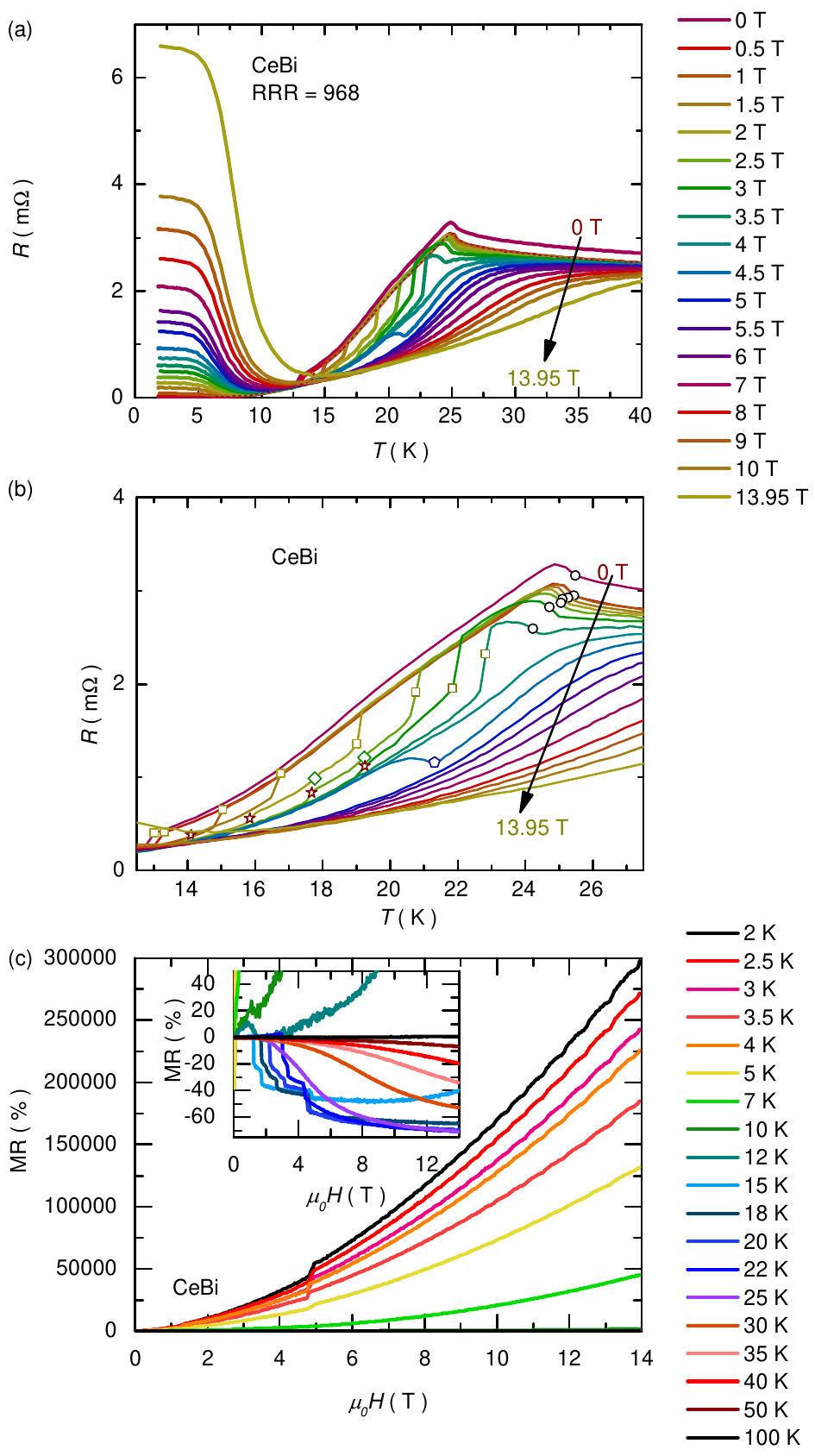}
\caption{(a) Resistance as a function of temperature measured at various applied fields from $0~$T to $13.95~$T, for $T\leq 40~$K. (b) A blow up of (a) with transitions marked with symbols same as in Fig. \ref{pd1}. $dR/dT$ is used as the criterion for evaluating the transition temperatures. (c) Magnetoresistance, $MR \% = \frac{R(H)-R(H=0)}{R(H=0)}\times 100$, as a function of applied magnetic field, measured at various temperatures ranging from $2~$K to $10~$K. Inset: MR data for $T\geq 12~$K showing negative values. Metamagnetic transitions are clearly visible as step-like features in the MR curves.}
\label{rtrh}
\end{figure}

The temperature dependent electrical resistance in zero field on a cleaved crystal of CeBi is shown in Fig. \ref{mtrt} (b). The resistance remains relatively invariant at high temperatures, followed by an $\sim20\%$ upturn and then a sharp decrease at $\sim25~$K, associated with loss of spin disorder. The RRR was calculated to be $968$. A similar behavior is observed for CeSb as well, but with a much less pronounced upturn \cite{Wiener2000}. The near 1000 RRR value attests to the high purity of the CeBi samples. This is further born out by the fact that (i) the magnetoresistance at low temperatures is huge and follows Kohler’s rule and (ii) Shubnikov-de Haas oscillation are detected. 
The inset of Fig. \ref{mtrt} (b) shows the low temperature region with the transitions clearly seen. The first one is around $25$ K, and another transition, observed as a relatively small jump in $R$ is seen around $12.5~$K. The transition temperatures are determined by taking the local maxima of the derivative $dR/dT$ \cite{Fisher1968,Ribeiro2003}.  

Figure \ref{rtrh} shows the temperature and magnetic field dependence of electrical resistance. The temperature dependence of $R$ with various applied fields is shown in Fig. \ref{rtrh} (a) for $T\leq40~$K. Figure \ref{rtrh} (b) shows the intermediate temperature regime where the various transitions are seen. Here, we have denoted the transition temperatures with open symbols (circles, squares, diamonds, and stars) with same shape and color, as those used for the corresponding transitions from magnetization data, in the phase diagram. When magnetic fields are applied, the first transition, denoted by black open circles in Fig. \ref{rtrh} (b), shifts to lower temperatures, and the second transition, denoted by dark yellow open squares in Fig. \ref{rtrh} (b), move to higher temperatures before merging into one, at around $4.5~$T. Between $2$ and $4~$T, two other anomalies appear, as a small jump in $R(T)$ data, denoted by open stars and diamonds in Fig. \ref{rtrh} (b). Similarly, the metamagnetic transitions could be observed in $R(H)$ data as well. The evolution of various features from $R(T)$ and $R(H)$ data follows the behavior in magnetization data and agree well with the phase diagram in Fig. \ref{pd1}. As temperature decreases both Figs. \ref{rtrh} (a) and \ref{rtrh} (c) show that once spin disorder (or magnon) scattering from the Ce$^{3+}$ is suppressed by field or temperature, a growing and large positive magnetoresistance emerges.

Figure \ref{rtrh} (c) shows the magnetoresistance, defined as $MR \% = \frac{R(H)-R(H=0)}{R(H=0)}\times 100$, as a function of $H$. A sharp feature in $MR$ vs. $H$ is seen around $4.8~$T, at $2~$K, coinciding well with the feature obtained in $M(H)$ data. At low temperatures there is a power-law like behavior with an experimentally determined power $n=1.6$ for fitting MR at $2~$K, above $6~$T with $MR=aH^n$. At $2~$K and $13.95~$T, MR reaches a value of $2.9\times 10^5\%$. Shubnikov-de Haas (SdH) oscillations are also observed at low temperatures and high fields. With increasing $T$ the quantum oscillations die away and the magnitude of MR decreases, as expected. Above $12~$K, a negative MR regime is observed. The step-like features in $MR$ vs. $H$ curves are associated with the metamagnetic transitions, as shown in the inset of Fig. \ref{rtrh} (c). At even higher $T$, above $100~$K, MR becomes positive again, but with comparatively small values. For instance, MR at $100~$K and $13.95~$T is $0.84\%$. 

\section{Discussion}

\begin{figure}
\centering
\includegraphics[scale=0.45]{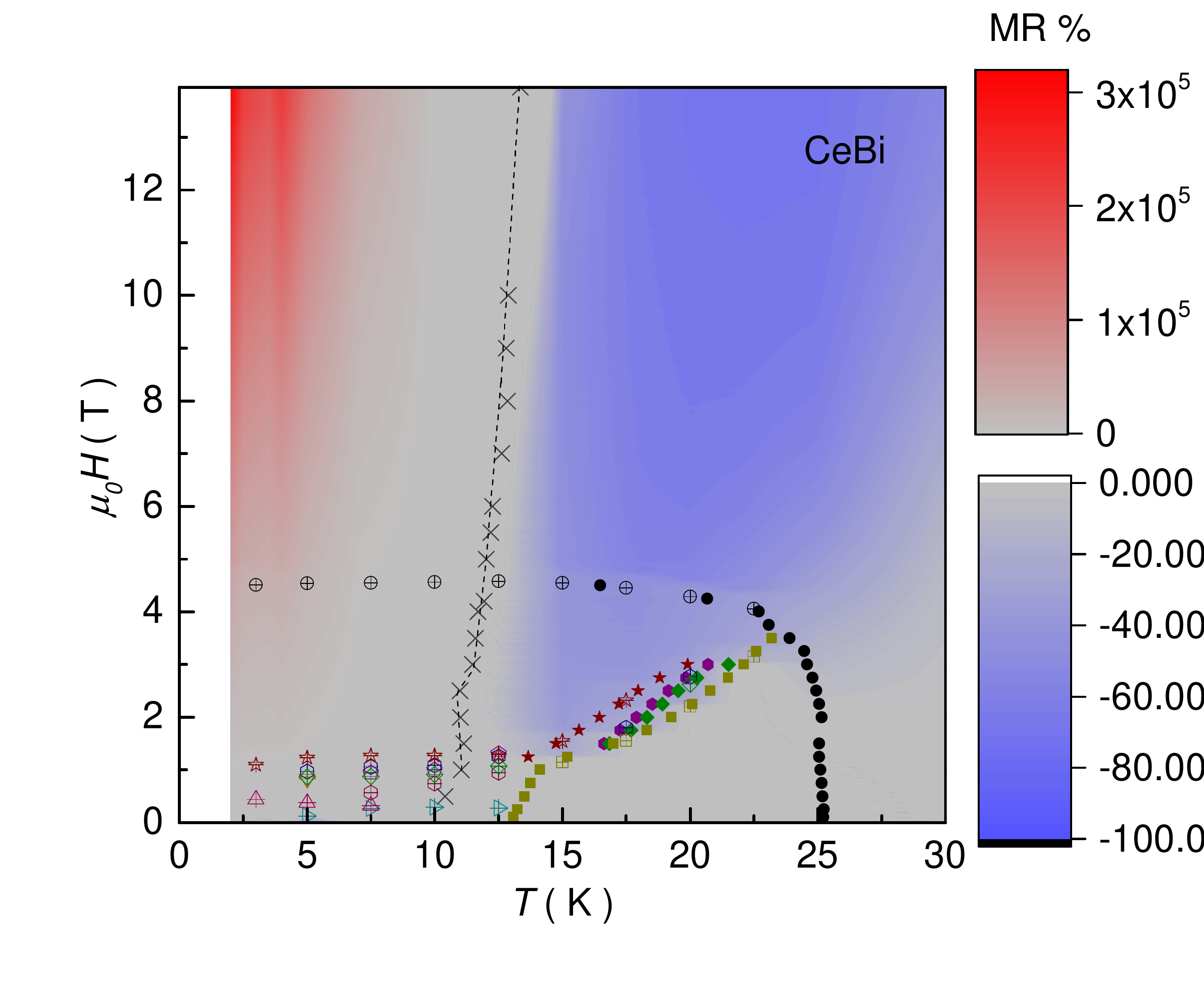}
\caption{MR plotted as a function of temperature and applied magnetic field as a false color map, overlaid with phase diagram showing various transitions in magnetization data. Blue region shows the negative MR regime and the transition through gray to red shows positive MR regions. In addition to the transitions shown in Fig. \ref{pd1}, the local minima from each $R(T)$ plot is marked by $\times$ symbol, which captures the change from negative to positive MR. Line connecting the crosses is guide to the eye.}
\label{mr}
\end{figure}

We can try to gain a better understanding of the magnetic and transport properties of CeBi by looking more closely at the $H-T$ phase diagram and comparing it with the magnetoresistance behavior. In the phase diagram shown in Fig. \ref{pd1}, one can observe an envelope of transitions, paramagnetic to antiferromagnetic at higher temperatures, and going to field polarized state at lower temperatures but with higher fields. Within this there is a clear and large region between $1.5$ to $4.5~$T at lower temperatures and reducing in width at higher temperatures above $15~$K. This corresponds to a regime with $M=1/2 \mu_{sat}$. This is seen as a clear plateau in $M-T-H$ data in Fig. \ref{mth}. Additionally, we have an antiferromagnetic region, existing between the transition at $25~$K and the lower one (near $\sim13~$K in lower fields), denoted by yellow squares in the phase diagram in Fig. \ref{pd1}. Then there are narrow stretches of magnetic phases existing between yellow squares and red stars in Fig. \ref{pd1}. The larger two of these regions correspond to phases with a net magnetization of $M=1/4 \mu_{sat}$ and $M=1/3\mu_{sat}$, where $\mu_{sat}$ is the saturated magnetization close to $2.14~\mu_B/f. u. $. There exists a third phase in between these two, and that has not been reported earlier. Once we enter the low temperature, low field regime, below $12.5~$K and less than $1.5~$T, we have multiple closely placed transitions and phases with ill defined boundaries. This likely corresponds to existence of many near degenerate states, as evidenced by the highly hysteretic behavior of both $M(T)$ and $M(H)$.

We can compare our $H-T$ phase diagram with the various MR regions, to better understand how magnetoresistance is being affected by the magnetic ordering. This is achieved through a false color plot of $MR$ as a function of $T$ and $H$, as shown in Fig. \ref{mr}. A change from blue to red colored region shows the variation from negative to positive large MR. 

The most conspicuous feature in the false color plot is the change from positive to negative MR. Although the lowest field data could suggest that this could be related to the transitions near $10 -15~$K, the fact that this MR sign change exists to fields 2.5 times larger than the metamagnetic transition to the saturated paramagnetic state indicates that this is not the case. The sign change in MR persists from the high field to low field region, unchanged by crossing the $\mu_0H\sim4.5~$T line. Based on the simpler, higher field $R(T)$ curves, the sign change appears to be associated with a change from a lower temperature, positive MR associated with minimal scattering from the already saturated Ce$^{3+}$ moments to a higher temperature, negative MR that is associated with the increasing applied field suppressing scattering from the Ce$^{3+}$ moments.  The negative MR region in the ordered part of the $H-T$ diagram indicates the region where magnetic scattering is most readily suppressed by increased field. One can not rule out the possible existence of a Lifshitz transition unaffected by the magnetic transitions, which could also cause such a change, but this would be coincidental. In other words, whereas electronic structural changes in zero field have been observed associated with magnetic transitions, both in optical and ARPES measurements \cite{Kimura2004, Li2019, Oinuma2019}, it is unlikely that this causes such a cross-over in MR behavior, in high fields, where we are in a saturated paramagnetic regime. This argument is further strengthened by the Kohler's rule analysis in the following paragraph.

\begin{figure}
\centering
\includegraphics[scale=0.55]{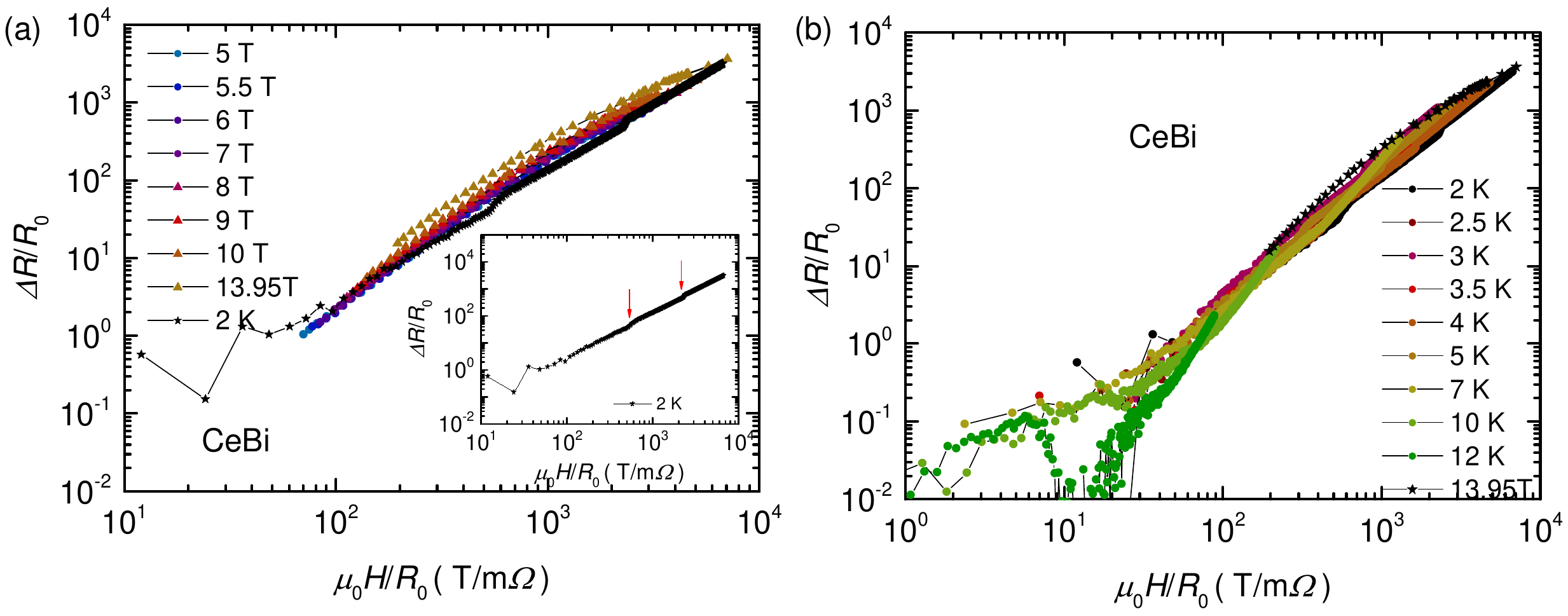}
\caption{(a) Kohler's plot obtained from $R(T)$ data taken at various fields above $5~$T, in the temperature range $1.8~$K$\leq T \leq 10~$K. In addition, MR data at $2~$K is plotted in the same way, which falls on top of the curves from $R(T)$ data. Inset: Kohler's plot for MR at $2~$K plotted separately. Two small features corresponding to the metamagnetic transitions are marked by red arrows. (b) Kohler's plot obtained from $R(H)$ data taken at various temperatures from $2~$K to $12~$K. The breakdown at $12~$K as we approach the magnetic transition is clearly seen.}
\label{kohler}
\end{figure}

The existence of a power-law like behavior of $MR(H)$ at low temperatures calls for the Kohler's rule analysis of this data set. Kohler's rule provides a simplistic approach, wherein the classical electron motion in an applied magnetic field leads to a scaling behavior of the form:
\begin{equation}
\frac{\Delta R}{R_0}=F(\frac{H}{R_0})
\end{equation}
Here, $F(x)$ is the scaling function and $\Delta R=R(T,H)-R_0$ where $R_0=R(T,H=0)$. Kohler's rule holds as long as there is a single, dominant scattering mechanism. Given that we think scattering from the Ce$^{3+}$ moment is minimal to the left of crossed line in Fig. 8, i.e. in the positive MR region, we can analyze our data in that region of the $H-T$ space. Figure \ref{kohler} (a) shows the Kohler's plot obtained from the resistance $R(T)$ measured at various fields above $5~$T, in the temperature range $1.8~$K$\leq T \leq 10~$K. Figure \ref{kohler} (b) shows the Kohler's plot from different $R(H)$ curves measured at various temperatures up to $12~$K. All the curves, except $R(H)$ at $12~$K, fall approximately on top of each other signifying the scaling behavior and its breaking as we approach the magnetic transition. In Fig. \ref{kohler} (a), one can see, the various $R(T)$ data roughly fall on top of each other, indicating the MR behavior at low temperatures being governed by the same physics across a wide range of applied fields. This also shows that the various magnetic transitions in the low temperature low field regime play a less dominant role in electronic transport, as the effects of these on electronic scattering are small (on the logarithmic scale) compared to the Kohler's rule behavior due to very high positive non-saturating magnetoresistance values observed at low temperatures. This can be emphasized more by plotting $\Delta R/R_0$ vs. $H/R_0$ using $MR(H)$ data at $2~$K. It agrees well with the curves from various $R(T)$ data as shown in Fig. \ref{kohler} (a). But additionally, if we look carefully at the $2~$K curve, shown separately in the inset of Fig. \ref{kohler} (a), one can see small glitches in it (indicated by red arrows in the figure), which corresponds to the field values of $1.1~$T and $4.8~$T, which are close to the fields where we observe the magnetic transitions. This clearly shows that the effect of magnetic transitions on the MR behavior in this regime is comparatively small. The slope of data shown in inset of Fig.\ref{kohler} (a) is roughly $5/3$.  Although the manifolds shown in the main body of Figs. \ref{kohler} (a) and (b) have some spread, they are consistent with this value as well.

Thus we can say, MR sign change tracks minimum in $R(T)$ close to $12~$K and is most clearly associated with a crossover from low temperature Kohler's rule - like behavior in $R$ with $H$ associated with a normal metal with an anomalously large MR to a higher temperature decrease in $R$ with $H$ associated with saturating the Ce spins and decreasing spin disorder/ magnon scattering.

The anisotropic MR behavior of CeBi was studied recently \cite{Lyu2019}, and it suggests a magnetization governed MR in the temperature regime $\sim 12.8$~K - $25~$K. This is not inconsistent with our results. Whereas, they observe a magnetization dependent magnetoresistance in the temperature regime between $T_N/2<T<T_N$, we focus on the Kohler's rule behavior below that and the shift to negative MR above that temperature region. 

\section{Conclusion}
We measured the magnetic and the transport properties of CeBi, on flux grown single crystals. From the magnetization data, we were able to construct a field-temperature phase diagram and identify regions with near degenerate states, as well as those with well defined magnetization values. We were also able to identify a new phase region in addition to the ones existing in earlier reports. In addition, we observed a non-monotonic behavior of magnetoresistance. A large MR was observed in the low temperature regime, where it has a power-law, non-saturated behavior, which obeys the Kohler's scaling rule. This gives way to the onset of a negative magnetoresistance region with increasing temperatures, when the magnetic scattering plays the dominant role. 

\section*{Acknowledgements}
Work at the Ames Laboratory was supported by the U.S. Department of Energy, Office of Science, Basic Energy Sciences, Materials Sciences and Engineering Division. The Ames Laboratory is operated for the U.S. Department of Energy by Iowa State University under
Contract No. DEAC0207CH11358. B. K.  was supported by the Center for the Advancement of Topological Semimetals, an Energy Frontier Research Center funded by the U.S. DOE, Office of Basic Energy Sciences. N. H. J. was supported by the Gordon and Betty Moore Foundation EPiQS Initiative (Grant No. GBMF4411). L.X. was supported, in part, by the W. M. Keck Foundation and the Gordon and Betty Moore Foundations EPiQS Initiative through Grant GBMF4411.


\end{document}